# Low Threshold Results and Limits from the DRIFT Directional Dark Matter Detector

J.B.R. Battat,[1] A.C. Ezeribe,[2,6] J.-L. Gauvreau,[3] J. L. Harton,[4] R. Lafler,[5] E. Law,[3] E.R. Lee,[5] D. Loomba,[5] A. Lumnah,[3] E.H. Miller,[5] A. Monte,[3] F. Mouton,[2] S.M. Paling,[6] N.S. Phan,[5] M. Robinson,[2] S.W. Sadler,[2] A. Scarff,[2] F.G. Schuckman II,[4] D.P. Snowden-Ifft,[3*] N.J.C. Spooner,[2] and N. Waldram[3]

*Corresponding author (ifft@oxy.edu)

(DRIFT collaboration)

[1]Department of Physics, Wellesley College, 106 Central Street, Wellesley, MA 02481, U.S.A.

[2]Department of Physics and Astronomy, University of Sheffield, S3 7RH, U.K.

[3]Department of Physics, Occidental College, Los Angeles, CA 90041, U.S.A.

[4]Department of Physics, Colorado State University, Fort Collins, CO 80523-1875, U.S.A.

[5]Department of Physics and Astronomy, University of New Mexico, NM 87131, U.S.A.

[6]STFC Boulby Underground Science Facility, Boulby Mine, Cleveland, TS13 4UZ, U.K.

We present results from a 54.7 live-day shielded run of the DRIFT-IId detector, the world's most sensitive, directional, dark matter detector. Several improvements were made relative to our previous work including a lower threshold for detection, a more robust analysis and a tenfold improvement in our gamma rejection factor. After analysis, no events remain in our fiducial region leading to an exclusion curve for spin-dependent WIMP-proton interactions which reaches 0.28 pb at 100 GeV/$c^2$, a fourfold improvement on our previous work. We also present results from a 45.4 live-day unshielded run of the DRIFT-IId detector during which 14 nuclear recoil-like events were observed. We demonstrate that the observed nuclear recoil rate of 0.31±0.08 events per day is consistent with detection of ambient, fast neutrons emanating from the walls of the Boulby Underground Science Facility.

## 1. Introduction

There is strong evidence from a variety of sources to suggest that 85% of the matter in the Universe is in the form of dark matter [1]. One possibility favored by theories beyond the Standard Model of particle physics is that dark matter consists of Weakly Interacting Massive Particles (WIMPs) [2]. As such, a large, international effort has been underway for decades to search for the rare, low-energy recoil events produced by WIMP interactions [1]. The DAMA collaboration measures an annual modulation in their event rate that they interpret as a WIMP detection [3]. However, other experimental results are in tension with this claim [4-8]. The primary goal of directional dark matter detectors is to provide a 'smoking gun' signature of dark matter [9, 10]. Such experiments seek to measure not only the energy, but also the direction of WIMP-induced nuclear recoils, thereby confirming their signals as galactic in origin. Numerous studies have shown the power of a directional signal, e.g. [11, 12]. Instead of order $10^4$ events required for confirmation via the annual modulation signature, only of order 10-100 events are required with a directional signature [13], assuming zero background. Additionally, instead of the easily mimicked annual modulation, the directional signal is fixed to the galactic coordinate system and is therefore less prone to false-positive detections. In recent years, several ideas for directional detection technologies have been proposed and revived [14-19]. At present, however, the only demonstrated, directional, detection technology to be deployed is recoil tracking in low-pressure gas time projection chambers (TPCs) [9].

The Directional Recoil Identification From Tracks (DRIFT) collaboration pioneered the use of low-pressure gas TPCs to search for this directional signal [20]. DRIFT utilizes negative ions (in particular, $CS_2^-$) to transport the ionization to the readout planes with diffusion at the thermal limit [20]. Background-free operation was made possible thanks to a unique thin-film cathode [22] and by the addition of 1 Torr of $O_2$ to the nominal 30+10 Torr $CS_2+CF_4$ DRIFT gas [23]. The addition of $O_2$ produces several species of so-called "minority carriers" that drift with slightly different velocities relative to the single species observed with the regular gas mixture [24]. The arrival time differences between these species are proportional to the distance from the readout plane, enabling a measurement of the distance, *d*, from the readout plane to the ionizing event. This measurement allows for the removal of all nuclear recoil backgrounds due to radioactive decays near the central cathode or readout planes. The detector fiducialization in the drift direction preserves a large nuclear recoil efficiency and directionality [25].

## 2. The DRIFT-IId Detector

DRIFT-IId is a dark matter detector designed to measure the ionization and

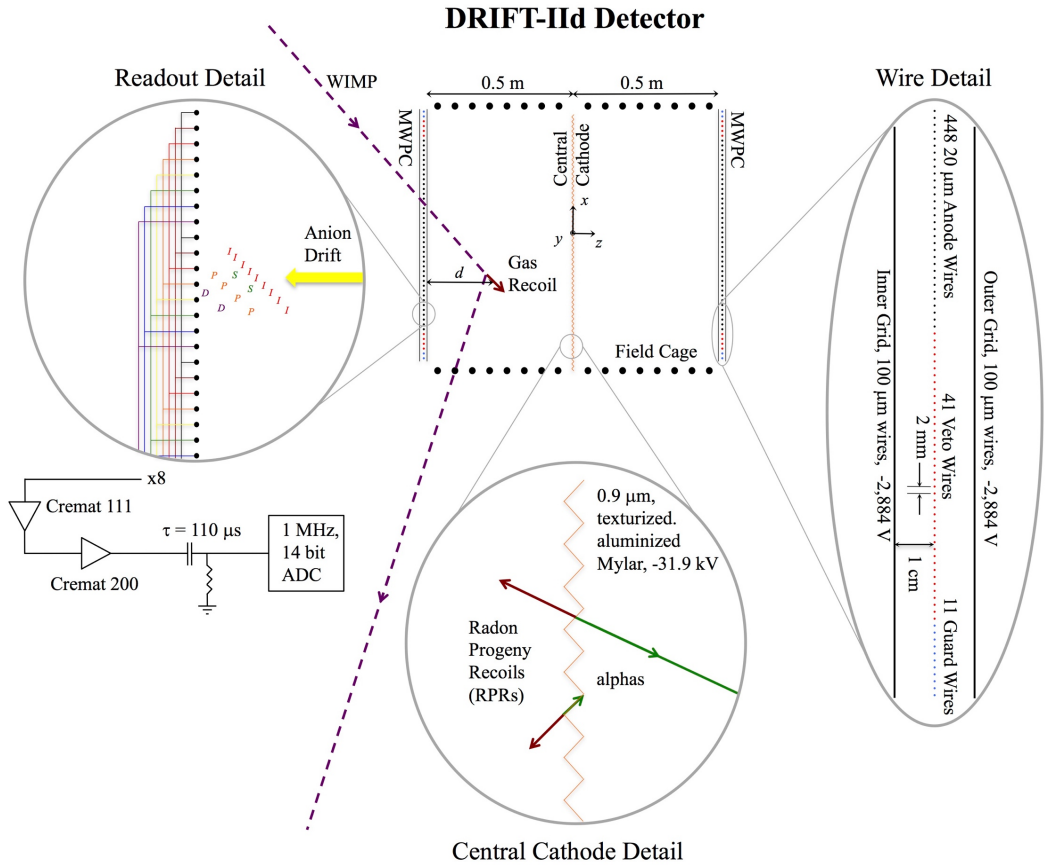

Fig. 1 – The DRIFT-IId Detector Detail shows a schematic of the detector, as viewed from above. The detector is composed of 2 MWPCs, a central cathode and a field cage. A WIMP is shown interacting in the gas inside the fiducial volume of the left side of the detector. The Readout Detail shows the separation of the minority carriers, labeled D, P, S on the way to the readout allowing the distance *d* from the MWPC to be determined. The 448 wires of the detector are grouped into 8 readout lines and read out as shown and discussed in the text. The Wire Detail shows details of the MWPCs. The Central Cathode Detail shows, schematically, the effect of the texturized central cathode on radon progeny recoil (RPR) events occurring there. For most alpha decays, see top event, ionization occurs on both sides allowing the event to be vetoed. In extremely rare cases, see bottom event, an alpha loses all of its energy in the cathode and cannot be vetoed.

orientation of ~keV/amu nuclear recoils produced by WIMP dark matter within its fiducial volume. As shown in Fig. 1, DRIFT-IId consists of two back-to-back TPCs separated by a cathode and read out by MWPCs on either side. The gas pressure is low to allow the low-energy recoils to produce tracks that are long enough so that the track shape is not lost to diffusion of the ionization as it drifts to the MWPCs. To limit diffusion, the DRIFT collaboration uses gas mixtures with electronegative $CS_2$ whose anions, unlike electrons, remain at room temperature despite drifting in electric field to gas density ratios,

E/N, in excess of $10^{-16}$ V cm$^2$ [21]. Additionally, DRIFT-IId utilizes $O_2$ in gas mixtures with $CS_2$ to enable fiducialization of the events along the drift direction [23]. A significant concentration of $CF_4$ allows DRIFT to search for spin-dependent (SD) WIMP-proton interactions via interactions with F whose nuclear spin is ½.

A brief description of the detector is provided here for convenience. Starting in the center, a unique 0.9 μm thick, texturized aluminized-Mylar cathode separates the left and right fiducial volumes of the DRIFT-IId detector, see Fig. 1 [22]. For all of the runs described below this thin film cathode was held at a voltage of -31.9 kV. The inner grid planes of the MWPCs are located 0.5m away from the central cathode (in the $z$ dimension, see Fig. 1) and are biased at -2.884 kV. As described below only the inner 0.803 m$^2$ of the MWPCs are read out for event signals, the remainder being used for 96 vetoes and field stabilization. 31 stainless steel tubes with 6 mm diameter surround each 0.401 m$^3$ fiducial region with stepped voltages to provide a uniform drift field. In this way two fiducial regions with a combined volume of 0.803 m$^3$ are created, each with a uniform 580 V/cm drift field. The entire vacuum vessel, described below, including the fiducial volume is filled with a mixture of 30+10+1 Torr $CS_2+CF_4+O_2$ providing 140(34.2) g of target(F) nuclei to search for WIMP dark matter.

The MWPCs, see Fig. 1 – Wire Detail, are made up of a central, grounded, anode plane of 552 20-μm-diameter stainless steel wires on a 2 mm pitch (measuring the $x$ extent of the events), sandwiched between two perpendicular, grid planes of 552 100-μm-diameter wires at -2.884 kV, again on a 2 mm pitch (measuring, using induced pulses, the $y$ extent of the events) and separated by 1 cm from the anode plane. 448 anode and grid wires form the lateral ($xy$) dimensions of the fiducial region giving a square, fiducial area of 0.803 m$^2$. 41(52) of the remaining wires form an anode(grid) veto on each side of this fiducial area. The balance, 11 wires on each side of the anode plane, are "guard wires" held at stepped voltages to prevent breakdown at the edges.

For both anode and grid, every 8$^{th}$ wire in the fiducial area (448 wires) is grouped together, providing 16 mm of readout per event in both $x$ and $y$, see Fig. 1 - Readout Detail. This is sufficient to contain the ~few mm WIMP recoils of interest. All signals are pre-amplified inside the vacuum vessel by Cremat CR-111 preamplifiers, then amplified by Cremat CR-200 shaping amplifiers (4 μs shaping time) outside of the vacuum vessel. Finally, the signals pass through a high-pass filter with time constant 110 μs and are digitized by 14-bit National Instruments PXI-6133 ADCs at sampling rate of 1 MHz with 0.152 mV resolution. The DAQ is triggered to read out all channels when any one of the anode signals rises above a threshold of 15 mV. This trigger level is half that used during the previous DRIFT-IId WIMP search [23] and provides a significantly higher sensitivity to nuclear recoil events in the analysis presented here. This is accomplished, without undue trigger rate, by box-car-smoothing all anode signals over 18 μs in software prior to software triggering. Both pre- and post-trigger data are recorded (3 ms and 10ms respectively). This is sufficient to contain the minority peaks which, typically, arrive pre-trigger and alphas which extend many milliseconds post-trigger. Anode and grid veto signals are read out separately for each MWPC. With the grouping described above, only 36 channels are needed to read out the entire detector.

Moving outward, a 6 mm thick acrylic shield prevents discharge to the grounded vacuum vessel. Two solenoid-activated Fe-55 sources irradiate, sequentially the left and right volumes of the detector with 5.9 keV X-rays to calibrate the gas gain. A stainless-steel vacuum vessel with interior volume of 1.5 m by 1.5 m by 1.5 m surrounds the entire detector.

To prevent neutrons from the walls of the cavern (rock-neutrons) from interacting in the DRIFT-IId detector, neutron shielding surrounds the vacuum vessel. The bulk of the shielding is composed of loose polypropylene pellets, with average density of 0.62 g/cm$^3$, contained in a wooden structure. The vacuum vessel is raised above floor level on three adjustable steel legs with pellets filling the gap below. The remainder of the shielding is under the floor in a cavity 10 m by 3.3 m by 45–55 cm deep. This shielding structure provides an average of 44 g/cm$^2$ polypropylene in all directions.

Finally, a custom-built gas system mixes evaporated $CS_2$ with a 90%-10% mixture of $CF_4$+$O_2$ to provide the requisite 30+10+1 Torr $CS_2$+$CF_4$+$O_2$ to the vacuum vessel. After flowing through the vacuum vessel, the bulk of the $CS_2$ is captured by pumping it to the bottom of a stainless-steel waste canister where it liquefies under several centimeters of water. Any remaining $CS_2$ is captured in a carbon trap. A web-based control system enables remote control of the detector and also provides system feedback on a number of channels (pressure, voltage, current etc.) at a rate of 1 sample every 4 seconds. The entire experiment is housed in the Boulby Underground Science Facility at a depth of 1.1 km. Further details of the DRIFT-IId detector can be found here [26].

3. Data

The data discussed in this paper were taken under a variety of experimental configurations and with a variety of radioactive sources for calibration.

*Shielded* – The purpose of the pellet shielding, described above, is to prevent rock-neutrons from entering the fiducial volume of DRIFT-IId. In this mode, we expect backgrounds to be suppressed to the fullest extent [27] and therefore it is these data which are most useful for setting limits on WIMPs. Data were collected with all of the shielding in place for 54.7 live-days (or 7.66 kg-days total or 1.87 kg-days F).

*Unshielded* – During these runs the above-ground pellets, described above, were removed; the under-floor shielding remained. The purpose of these runs was to attempt to measure the very low recoil rate from rock-neutrons [27]. This measurement is crucial for understanding the neutron background in Boulby. 24.3 live-days of unshielded data were collected prior to the start of the shielded runs and 21.1 live-days of unshielded data were collected after the end of the shielded runs. Thus, a total of 45.4 live-days of unshielded data were collected and bracket the shielded data. From the first unshielded run to the last unshielded run 402 days elapsed encompassing all of the runs discussed here.

*Fe-55 Calibration* – Regardless of the type of run, data taking is halted every 6 hours and Fe-55 calibration data is acquired. Shutters in front of the Fe-55 sources are opened (one at a time, left and right) exposing the fiducial volume to 5.9 keV X-rays. As discussed in [28] the W-value for 30+10 Torr $CS_2+CF_4$ is 25.2±0.6 eV. 5.9 keV Fe-55 X-rays therefore produce events with 234 *NIPs* (number of ion pairs). As in [23] we assume that the addition of 1 Torr of $O_2$ does not significantly affect the W-value of the 30+10+1 Torr $CS_2+CF_4+O_2$ gas mixture. Measuring the ionization collected on the wires allows DRIFT to calculate the gas gain and as a result monitor gas quality. In a typical Fe-55 run several thousand events are collected and analyzed.

*Cf-252 Calibration* – Cf-252 neutrons produce S recoils with an energy spectrum nearly identical to the recoils from massive WIMPs [29]. As such it is an ideal source for calibrating and monitoring our recoil detection efficiency. These data are crucial to the recoil analysis described below. During a Cf-252 calibration run the center of the source is placed at a standard source location 10 cm above the vacuum vessel and centered on the central cathode, (43 cm, 90 cm, 0 cm) relative to the coordinate system shown in Fig. 1. This position is inside the pellet shielding for the shielded runs. For these runs, pellets inside a plastic source-tube were removed, the source inserted and neutron recoil data taking commenced. Over the course of the runs described and analyzed in this paper, Cf-252 exposures were done once every 12 days, on average. In the middle of the first unshielded run a new Cf-252 neutron source replaced the old Cf-252 neutron source described and calibrated in [27]. As discussed below the new source was calibrated using the DRIFT-IId detector and found to be 27.2±1.2 times stronger than the old source. This, along with the previous calibration of the old source [27], allows us to calculate the strength of the new source. The activity of the new source in the middle of the shielded runs was 28,000±2,000 neutrons/s.

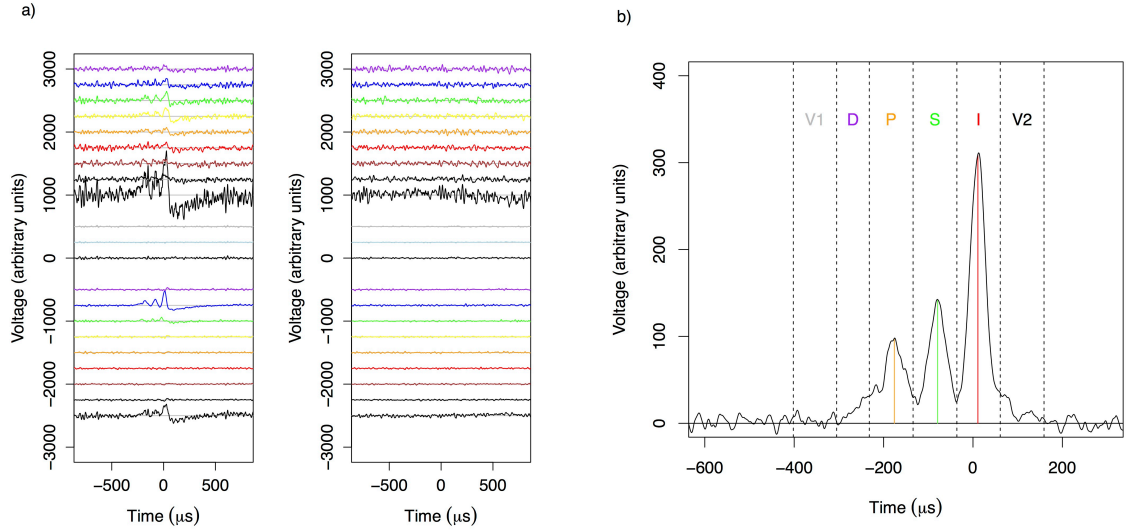

Fig. 2 – a) A typical neutron recoil event which generated 2,650 *NIPs* (104 keVr F). All 36 readout lines are shown separated from one and another by offsets purely for visualization purposes. On the bottom left (purple to black) are the 8 anode lines from the left detector and the anode sum line. As can be seen an event occurred primarily over the blue colored line at around $t = 0$ μs. The minority carriers (small peaks) arrived ahead of the main carriers. For this event the time separation of the peaks indicates that this event occurred 21.6 cm away from the detector. On the top left are the 8 grid lines of the left detector showing a similar time structure for the event, as expected and the grid sum line. Directly below the grid sum line are the grid (gray) and anode (light blue) veto lines. The sum veto line is below them. As can be seen there is no signal on the veto line. The right lines show nothing happening for this event on the right side of the detector. b) This plot shows the peak fitting results for the biggest amplitude line for the event, the blue line on the anode from a). The overshoot is removed in analysis and a peak-search algorithm, described in the text, divides the event into 6 bins V1, D, P, S, I, V2. The letters D, P, S and I designate different types of anions [24]. Nominally there should be no ionization in the veto bins (V1 and V2) but because the I peak is so large its tail often produces some ionization in the V2 bin. The location of the I-peak is shown in red while the locations of the S and P peaks are shown in green and orange. The D peak is so small that its location is not calculated or used in the analysis.

*Co-60 Exposures* – As discussed in [27] DRIFT's gamma rejection ability was $<3 \times 10^{-6}$ between 1,000–6,000 *NIPs* based on a Co-60 exposure though of limited exposure. To improve this limit three Co-60 sources with a total decay rate, at the time of the exposure, of 56 kBq were placed in the source-tube, inside the shielding and 2.90 live-days of data taken.

## 4. Data Reduction

The goal of any DRIFT data reduction procedure is to select, with high efficiency, nuclear recoils events within the fiducial volume and to reject, with high efficiency, all other types of events. An example of a typical nuclear recoil event selected from a Cf-252 neutron run by the procedure described below is shown in Fig. 2a. For nuclear recoil events passing quality cuts, the critical output is a measurement of the ionization of the event, *NIPs*, and the distance away from the MPWC readouts, *d*.

As discussed in [20] the physics of ionizing radiation provides a natural way to distinguish nuclear-recoil events from Compton electron-recoil events. For similar ionization electron-recoil events are at least 10 times longer than nuclear-recoil events. Therefore, electron-recoil events have an ionization density at least 10 times smaller than nuclear-recoil events. To utilize this the shaping time of DRIFT electronics is designed to be short compared to the typical duration of arrival of ionization. The hardware trigger threshold for DRIFT-IId, discussed above, is best thought of as a threshold on *ionization current*. Therefore, a reasonable threshold for short-duration, nuclear-recoil events typically rejects many, long-duration electron-recoil events. For example, and discussed in more detail below, the rate of Co-60 Compton interactions in the fiducial volume during the Co-60 runs is predicted to be many tens of Hz whereas during these runs the detector trigger rate rose only ~1.5 Hz above typical ~1.5 Hz background rates. The hardware threshold was the first, Stage 0 cut, for this analysis rejecting many electron-recoil events.

The majority of the ~1.5 Hz of events triggering the detector during the background runs were due to MWPC events and alphas. MWPC events are created by ionization in the gas volume between the grids of the MWPCs. In the large electric field in this region even events producing low ionization can produce ionization currents large enough to trigger the DAQ. The small increase in trigger rate during the Co-60 runs was due to this effect. Alpha events are due to radon decay in and around the fiducial volume of the detector and radon progeny decays on the surfaces surrounding the fiducial volume. To reject both of these types of events Stage 1 cuts were applied to the data.

First, though, raw DRIFT waveforms, recorded to disk, have a variety of simple, harmonic backgrounds due to the high voltage power supplies (several tens of kHz) and line pickup (several tens of Hz). The former were dealt with by Fourier analyzing each line, notching out the unwanted high frequencies and then reconstructing the events. The latter were dealt with by fitting the waveforms to a series of 5 low frequency harmonics and subtracting the fits from the waveforms. Nuclear recoil events are several hundreds of μs long and therefore have a bandwidth between these noise extremes.

A region of interest (ROI) between -700 μs and +700 μs relative to the trigger time is sufficient to contain all nuclear recoil events of interest. A number of basic statistics were calculated for any line with an amplitude of 9.9 mV or bigger (a hit) within the ROI. These statistics are discussed in detail in [30] and enable the Stage 1 cuts.

To remove large, MWPC events any event which saturates the ADCs (very large ionization current) was removed. To catch smaller MWPC events any event with a risetime (defined as the time between 10% of full height and 90% full height) of <3 μs was removed. To remove alphas any events which extended beyond the ROI (30 cm long alphas can be up to 4.7 ms long), hit more than 8 anode wires (16 mm) or hit both sides of the detector were removed. Alphas often cross the vetoes. A 30 mV cut on the left and right summed veto line served to remove most alphas crossing the vetoes. Finally, any hit lines were required to be adjacent to each other, as expected for nuclear recoil events. The Stage 1 cuts reduced the number of events going to the next stage of the analysis to ~30% of the events on disk.

AC coupling in the shaping electronics causes the signals to undershoot the baseline as can be seen in Fig. 2a. Specifically, the presence of minority peaks and the time scale of the coupling produces a large undershoot at the main, I, peak. For the remaining analysis, it was imperative that this undershoot be corrected. Mimicking the shaper electronics, the overshoot correction algorithm utilized a two-stage procedure with each channel individually tuned. Fig. 2b shows the result of this procedure.

With the proper baseline restored, the first important output parameter, *NIPs*, was calculated. As discussed in [30] the integral of the voltage over time, called the event sum, is proportional to the *NIPs* associated with the event. In practice, the event sums from individual hit lines *and* adjacent lines were included to calculate this parameter. The proportionality constant between the event sum and the *NIPs* was obtained from Fe-55 data.

Because Fe-55 events have such small amplitudes, data were taken, without a threshold, continuously, during an Fe-55 run. After the noise reduction techniques discussed above were applied, the data were boxcar-smoothed over 50 μs. Then starting from $t = -2$ ms lines were searched for smoothed voltages above a threshold of 2.3 mV. When one was found the event statistics, including event sums, were generated as described above. The search then proceeded to the next line whose smoothed waveform was above threshold and the process repeated until $t = +9$ ms. For a typical Fe-55 run and analysis several thousand events were gathered in this way. Some simple cuts removed alpha and MWPC events. The remaining Fe-55 event sum data were fit to a Gaussian and the centroid located. This centroid was equivalent to 234 *NIPs,* as described above, and in this way we obtained the proportionality constant for that time. Fe-55 data, taken every 6 hours, over many days were fit to obtain a smoothed set of proportionality constants good to better than 1% for any event between Fe-55 calibrations. Table 1 shows the conversion from observed ionization into energy for C, F and S recoils [30].

Table 1 – Conversion of recoil energy to *NIPs* and *Range* for C, F and S.

| Recoil Energy (keVr) | C Ionization (*NIPs*) | C Range (mm) | F Ionization (*NIPs*) | F Range (mm) | S Ionization (*NIPs*) | S Range (mm) |
|---|---|---|---|---|---|---|
| 10 | 164 | 0.5 | 140 | 0.3 | 115 | 0.2 |
| 20 | 395 | 0.9 | 332 | 0.6 | 259 | 0.4 |
| 30 | 659 | 1.3 | 552 | 0.9 | 416 | 0.5 |
| 40 | 946 | 1.7 | 792 | 1.3 | 588 | 0.7 |
| 50 | 1243 | 2.1 | 1055 | 1.6 | 773 | 0.8 |
| 60 | 1559 | 2.5 | 1326 | 1.9 | 966 | 1.0 |
| 70 | 1877 | 2.8 | 1616 | 2.2 | 1167 | 1.2 |
| 80 | 2205 | 3.2 | 1911 | 2.5 | 1370 | 1.3 |
| 90 | 2547 | 3.5 | 2223 | 2.8 | 1575 | 1.5 |
| 100 | 2886 | 3.9 | 2528 | 3.1 | 1788 | 1.6 |

The most important Stage 2 cut was the *NIPsRatio* cut. The purpose of this cut was to ensure that all events exhibit ionization consistent with the presence of minority peaks. This cut had the additional benefit of cutting events near the MWPC, such as radon progeny recoils (RPRs) emitted from the inner grid wires, because, for these events, the minority peaks do not have time to separate from the main, I, peak. The *NIPsRatio* cut was implemented in the following way. Starting from the point of highest voltage on any hit line within the ROI, the time at which the voltage first achieved 25% of this maximum was located. The *NIPsRatio* is the ratio of the ionization before this time to the ionization after this time summed over all hit lines. Events with *NIPsRatio* > 0.3 were kept, the rest were cut. An additional harder cut on smoothed veto lines at 2.3 mV ensured all events were fully fiducialized. The Stage 2 cuts reduced the number of events going to the next stage of the analysis to ~3.5% of the events on disk.

The Stage 3 cuts ensured that events exhibit clear minority peaks so that the distance from a MWPC, *d*, could be determined accurately. The analysis algorithm was "trained" with the use of the neutron recoil data. The characteristics used for comparison were,

1) The ratio of the size (*NIPs*) of the minority peaks to the size of the I peak.

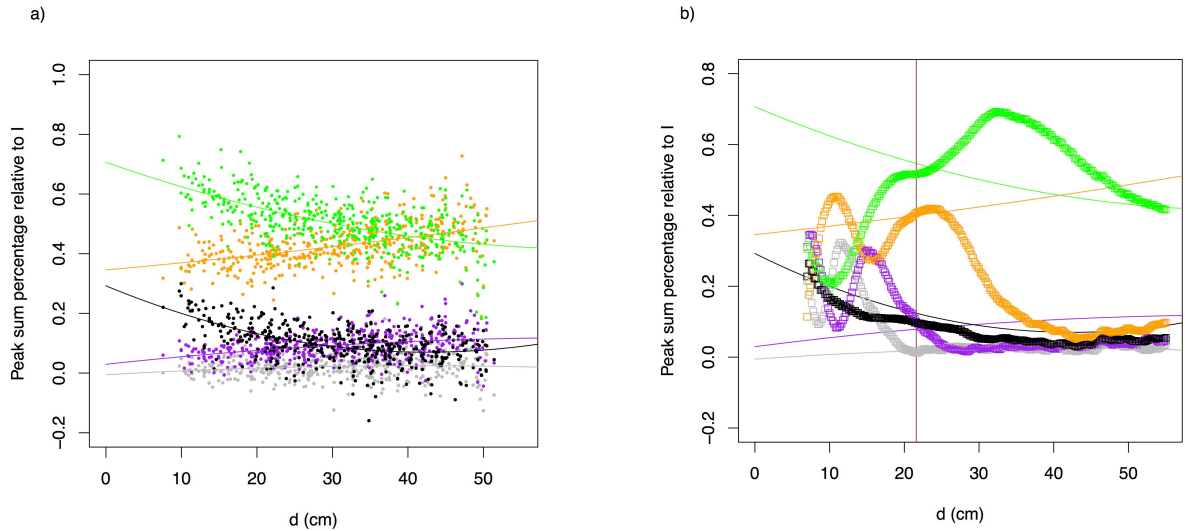

Fig. 3 – a) This figure shows size ratio data for several hundred neutron events. The vertical axis is the ratio of the ionization inside a particular bin (V1, S, P, D or V2) to the I bin. V1, S, P, D and V2 peak ratios are shown in gray, green, orange, purple and black respectively. The horizontal axis shows the manual measurement of *d* from the position of the P and I peaks. Fits to the data are shown. b) This plot shows the peak ratios, displayed as boxes, for various trial *d*s for the biggest amplitude (blue) line of the event shown in Fig. 2. For each trial the reduced $\chi^2$ between the trial values and the fits, shown in colored lines, is calculated. The minimum, reduced $\chi^2$ gives an approximate value for *d*, shown with a vertical brown line, for this event. At the selected *d* the data come close to matching their respective curves.

2) The ratio of the height of the minority peaks to the height of the I peak.

3) The temporal position of the minority peaks to the position of the I peak.

To assess any of this data the approximate positions of the peaks had to be determined. The approximate positioning of the peaks came from the size ratio analysis so the analysis of this ratio and the cuts on it were handled a little differently from the other two.

To obtain the size ratios and the approximate position of the peaks the following procedure was used. First a small sample of events from a neutron run were analyzed manually and the position of the P and I peaks located. Poor or random events were rejected. This procedure was developed because of the low signal to noise ratio after the Stage 0, 1 and 2 cuts with the old neutron source. With the new neutron source this first step could have been eliminated but, for sake of consistency, it was kept for the entire analysis. Knowledge of the approximate positions of the P and I peaks allowed the positions of the other peaks to be located. The event was then "binned" with spacing proportional to the P-I time difference as shown in Fig. 2b and the *NIPs* calculated for each peak. Fig 2b identifies two veto bins, V1 and V2, located before and after the event respectively. Nominally good events would show no ionization in these bins. The ratio of the *NIPs* in each of these bins to the *NIPs* in the I bin was then calculated and saved. 2$^{nd}$ order fits to the data and the spread as a function of *d* were then calculated, see Fig. 3a. It was observed that these fits varied, slightly, with time so non-neutron data, background data for instance, were compared to fits of neutron data closest to them in time.

All events passing the Stage 0, 1 and 2 cuts for all data close in time were then compared to these fits in the following way. As illustrated in Fig. 3b, for each event and for *d* trials ranging from 7 cm up to 55 cm with 0.1 cm increments a reduced chi-squared was calculated utilizing the RMS spread of the data shown in Fig. 3a. The *d* with the minimum reduced chi-squared was recorded as well as the minimum reduced chi-squared, $\chi^2_{size,min}/\upsilon$, where $\upsilon$ is the degrees of freedom, 4 in this case. A cut on $\chi^2_{size,min}/\upsilon$ at 4 insured that the bulk of the data was of high quality.

With an approximate value for *d* obtained from the above procedure each event was binned and, within each bin, the peak height obtained. The ratios of the heights of the P and S peaks to the height of the I peak was calculated for neutron data passing the Stage 0, 1 and 2 cuts and $\chi^2_{size,min}/\upsilon$ < 4. The D, V1 and V2 data were not used. Fits to neutron data and spreads for the height ratio data were calculated as a function of *d*. All data were then compared to these fits, using the approximate value for *d* and a reduced chi-squared for the height, $\chi^2_{height}/\upsilon$, data calculated.

Finally, within each bin the peak position was obtained. The differences in time between the positions of the D, P and S peaks to the position of the I peak were calculated from neutron data passing the Stage 0, 1 and 2 cuts and $\chi^2_{size,min}/\upsilon$ < 4. Fits to the data and spreads were calculated as a function of *d*. All data were then compared to these fits and a reduced chi-squared for the position data, $\chi^2_{position}/\upsilon$, calculated.

Only data with $\chi^2_{size,min}/\upsilon$ < 4, $\chi^2_{height}/\upsilon$ < 3 and $\chi^2_{position}/\upsilon$ < 3 passed the final, Stage 3, cuts.

The calculated time intervals between P and I peaks were found to produce the best estimate of *d*. As shown below, background data show a large concentration of events around the central cathode as expected from RPR events. The position of the central cathode is well known to be 50 cm away from the detectors with small error. Because of this, the average time difference for the P and I peaks for these runs was used to calibrate the conversion of the P-I time difference into distance from the detector. This time dependent calibration was found to vary by ~2% (1 cm) over the course of data taking and helped to improve our measurement of *d*. Data for *d* come from the P-I time difference obtained in this way.

Only ~0.003% of the events on disk pass all (Stage 0-3) cuts for a typical shielded data run.

## 5. Efficiency Map

Fig. 4 shows $d$ vs. *NIPs* for all of the shielded neutron runs. Only events with $700 <$ *NIPs* $< 6000$ were considered in this analysis. Furthermore because of interference from background RPR events, their spread on the central cathode and difficulties fitting events with small $d$ due to overlapping minority peaks, a fiducial region from $11.0$ cm $< d < 48.0$ cm was chosen. The fiducial window for this analysis is therefore ($700 <$ *NIPs* $< 6000$ and $11$ cm $< d < 48$ cm). This is shown in Fig. 4 with a large, lower, tan rectangle. Within this rectangle the data at small *NIPs* values show a border which has a positive slope. This is due to diffusion of events which decreases their amplitude below threshold, an effect that grows with $d$. Another boundary at small $d$ with positive slope within the fiducial region is also observed. This is caused by long, energetic recoils with a significant $z$ component such that the separation of the peaks is of order the length of the track. Because of the overlap, these events were cut by the *NIPsRatio* cut.

When the new neutron source was obtained two runs, old and new, were performed, back-to-back, in identical configurations and the rate of events appearing in the fiducial window calculated with identical cuts. The ratio of these rates allowed us to cross-calibrate the sources.

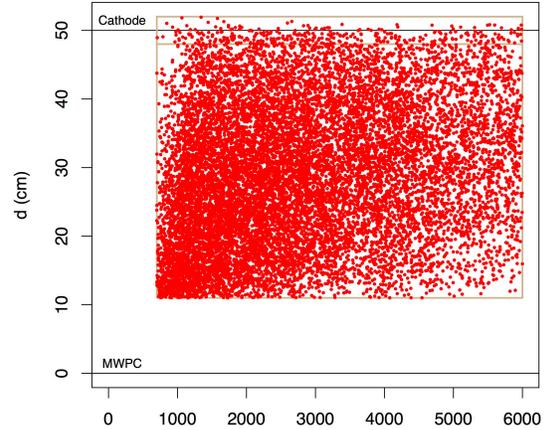

Fig. 4 – Plot of $d$ vs. *NIPs* for all accepted, shielded neutron events. 14,240 events occurred within the fiducial region of the detector, shown in the figure as a large, lower, tan rectangle.

As discussed in [23], an empirical approach has been adopted to deal with the difficult to model effects which produce these data. A detailed GEANT4 Monte Carlo (MC) [31] simulation of the DRIFT-IId detector was developed. The simulation reproduced the experimental setup where the Cf-252 source was enclosed in its protective Pb canister. The neutrons were fired isotropically with energy and frequency sampled from the energy spectrum produced using SOURCES-4C. Any nuclear recoils in the fiducial volume were recorded. To reduce the computational burden, we increased the pressure of the gas mixture by a factor of 25 (to 1,025 Torr), taking advantage of the linear relation between pressure and recoil rate. It was found that no, non-linear corrections were required since the probability of multiple scattering in the gas was negligible below 4,000 Torr [27]. 0.9 billion neutrons from a Cf-252 source were tracked equivalent to, using the measured strength of the neutron source, 9.4 days of live time, more than 10 times the live time of the actual neutron runs. From these data, the recoil type, energy and *d* could be easily extracted. Table 1 was then used to generate simulated *d* vs *NIPs* data. A small sample of the simulated recoil data is shown in Fig. 5a.

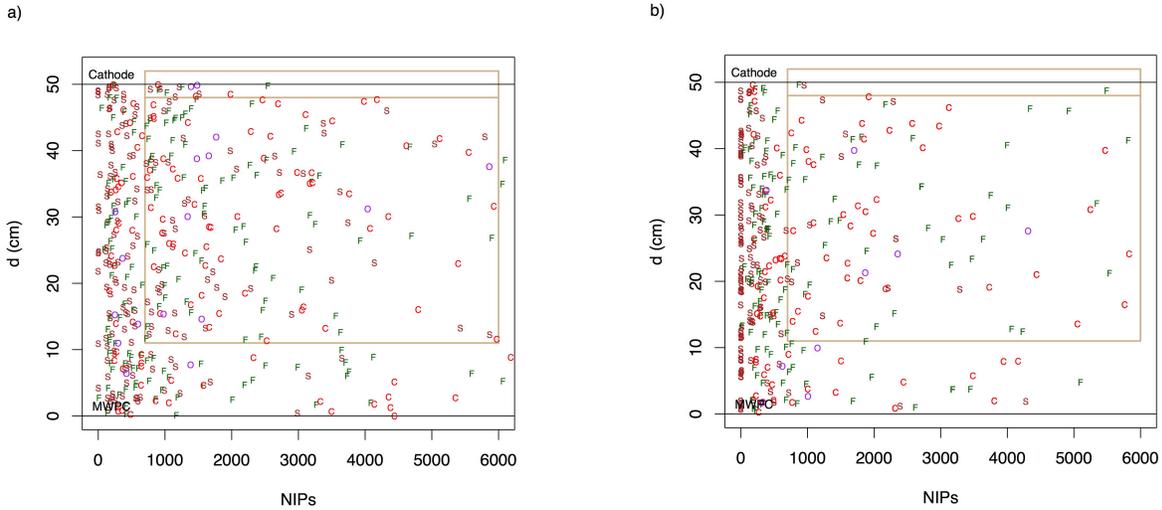

Fig. 5 – a) A small sample of Cf-252 neutron recoil events generated by GEANT4. The color and letters indicate the recoiling nucleus type. b) A sample of rock-neutron events generated by GEANT4. The concentration of events at low *NIPs* is due to multiple scattering of the neutrons in the walls of the Boulby cavern before entering the detector, see text for more details.

Real, Fig. 4, and simulated, Fig. 5a, data were binned within the fiducial window in *NIPs* and *d* producing cells. The number of events in each cell was counted and divided by the live time to produce maps of detected and predicted rates. The ratio of detected to predicted rates produces an efficiency map for the detector, shown in false color in Fig. 6. The advantage of this empirical approach is that all of the difficult-to-model physical and analysis effects inherent in events appearing on Fig. 4 can be easily incorporated into the detector model.

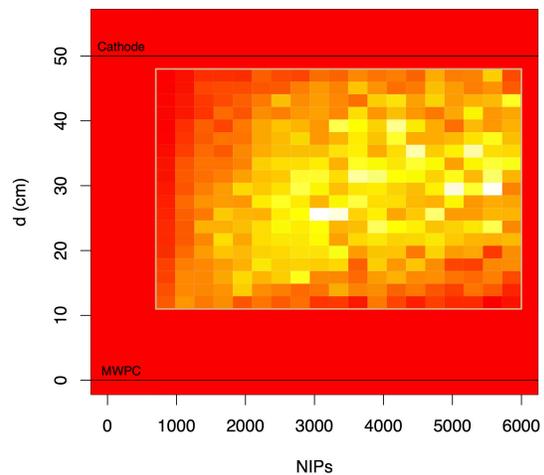

Fig. 6 – Efficiency map for the detector. White represents 100% detection efficiency while red represents 0% detection efficiency.

Because of the greater amount of neutron and GEANT4 data in this analysis the number of cells in the current map was increased by a factor of ~4 from the previous [23] map. The increased resolution is particularly important at low *NIPs* where the majority of dark matter recoils are predicted to occur. This map shows significantly increased probability for detection at lower *NIPs* values due to the smaller trigger threshold. Finally, unlike the previous efficiency map, no normalization of the MC live time was required.

This map can be used for predictive purposes. For any purported nuclear recoil signal the *NIPs* and *d* for all events expected in the live time for the run would need to be calculated. These events would then be binned equivalently to the efficiency map shown in Fig. 6. The number of events in each cell would then be multiplied by the efficiency for each cell and summed to obtain the predicted number of events that would have been detected. This number can then be directly compared to the observed number of events for the run.

## 5. Results

Fig. 7 shows the combined *d* vs. *NIPs* for data passing the cuts in 54.7 live days shielded running. The RMS spread of the events around the central cathode is 0.6 cm however a large contribution to this spread is not measurement but comes, instead, from wrinkles in the Mylar. No events were observed in the fiducial window of the detector for these shielded runs. These background-free data form the basis for the limits presented at the end of this paper.

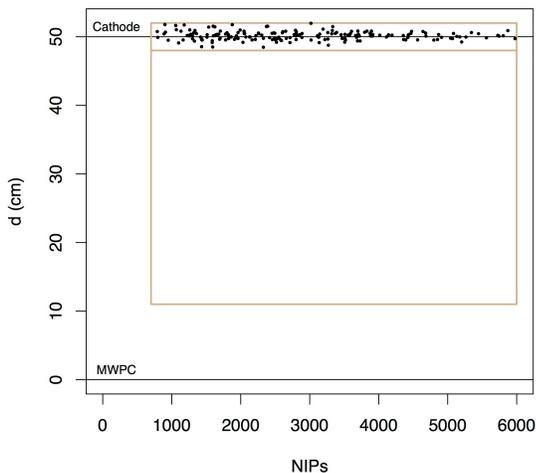

Fig. 7 – *d* vs. *NIPs* data for 54.7 live-days of shielded background data. All of the events passing the analysis cuts cluster around the central cathode consistent with the expectation of RPRs events there. In the fiducial window, large tan rectangle, no events were observed. This background-free result provides us with a limit on WIMP dark matter.

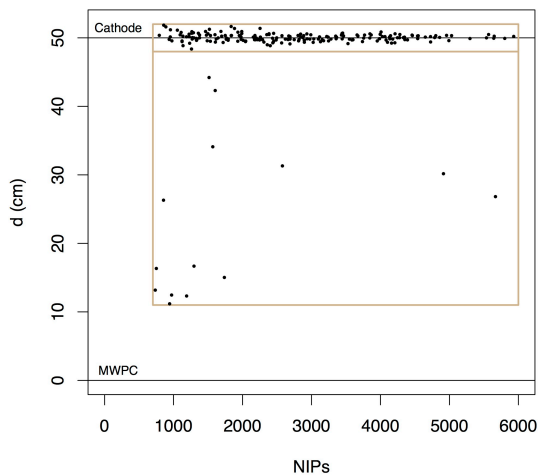

Fig. 8 – *d* vs. *NIPs* data for 45.4 live-days of unshielded background data. In contrast with the shielded background data, 14 events appeared in the fiducial window of the detector consistent with nuclear recoils created by rock-neutrons, see text.

Fig. 8 shows the combined $d$ vs. *NIPs* for data passing the cuts in 45.4 days of live time with the detector unshielded. In contrast with the shielded data 14 events appear in the fiducial window of the detector at an average rate of 0.31±0.08 events per day. The rate of events in the fiducial window before and after the shielded runs was consistent with limited statistics, 8 events before and 6 after.

## 6. Discussion of Unshielded Results

The unshielded detector was subject to a higher flux of ambient gammas produced in the walls of the underground lab, rock-gammas. To explore rock-gamma contamination we exposed the detector to three Co-60 sources as described above. After the standard analysis, no events were observed in the fiducial window.

To interpret this result a detailed GEANT4 MC simulation was performed. A 56 kBq Co-60 source was simulated inside its plastic container placed in the standard source location. Co-60 decays produce two isotropic gamma rays with 1.17 and 1.33 MeV respectively. Electron recoils in the fiducial volume were recorded with electron energies converted to *NIPs* using a conversion factor of W = 25.2 eV/*NIPs* [28]. For this simulation, self-shielding was observed within the gas so the simulation was performed at the nominal 41 Torr. Within the fiducial window GEANT4 predicted a rate of 46.26±0.07 Hz from the Co-60 sources. That none were observed improves our gamma rejection by more than a factor of 10, to $1.98\times10^{-7}$ (90% C.L.), from our previous result [27].

To establish what fraction of the 14 observed events could be due to Compton interactions from rock-gammas a simulation of rock-gammas was performed. As suggested by [32] the rock around the laboratory was homogeneously populated with U-238 (70 ppb), Th-232 (125 ppb) and K (1130 ppm) using GEANT4. For U-238 and Th-232, the whole decay chains were simulated assuming secular equilibrium. It was found that only the first 25 cm of rock contribute substantially to the gamma ray flux. To save computation time then, only the inner 25 cm of rock was populated with the aforementioned radionuclides. Electron recoils in the fiducial volume of the unshielded detector were recorded. Fig. 9 shows a comparison of the predicted *NIPs* spectra from each exposure. While not identical the spectra are remarkably similar. We assume this in applying the Co-60 rejection factor to rock-gamma events below. Within the fiducial window GEANT4 predicts a rate of 7.8±1.3 Hz from rock-gammas or $3.1\times10^{7}$ events in the 45.4 live-days of the unshielded exposure. Applying the rejection factor gives an upper limit (90% C.L.) on gamma contamination of 6 events. All 14 events appear well contained in *x*, *y* and *z*, inconsistent with a large contamination by rock-gamma events. More Co-60 exposures are planned to further, rigorously lower the upper limit on gamma contamination.

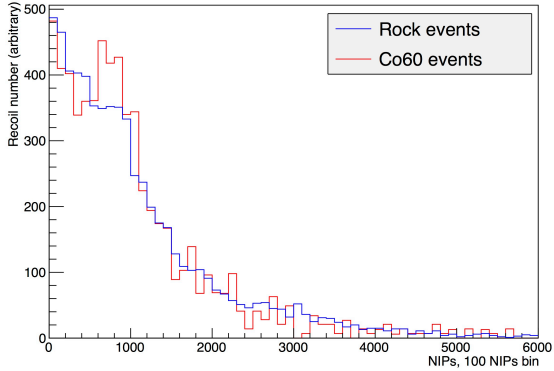

Fig. 9 – GEANT4 simulation of events created by gamma rays within the fiducial volume of the detector from a Co-60 source on top of the detector (red) and U-238, Th-232 and K-40 from the rock (blue) as described in the text. The spectra are largely identical making Co-60 a good stand-in for exposure to the ambient rock-gammas.

The observed recoil rate can be utilized to measure the U and Th content in the walls of the Boulby Lab. A GEANT4 simulation tracked $5.5\times10^8$ neutrons generated from U-238 (10 ppb) and Th-232 (10 ppb) in 3 m of rock surrounding the lab. Because of self-absorption, additional rock did not appreciably change the interaction rate. If any of these neutrons interacted in the fiducial volume of the detector, information on the recoil was recorded. Fig. 5b shows a sample of these rock-neutron events. This MC data was then binned in the usual way and multiplied by the efficiency map. The predicted detectable rates were 0.021±0.003 events per day for 10 ppb U-238 and 0.010±0.001 events per day for 10 ppb Th-232. Assuming all of the 14 events are neutron recoils, that Th-232 is two times as abundant as U-238 [33] and that both decay chains are in secular equilibrium, we find 77±20(stat)±7(sys) ppb U-238 and 150±40(stat)±10(sys) ppb Th-232. Table 2 summarizes previous measurements of these quantities by other groups.

The agreement shown in Table 2 between our results and other, more traditional, methods suggests that our assumption of zero gamma contamination was correct and motivates further improvements in the gamma rejection factor for DRIFT. If the rejection factor could be pushed down another order of magnitude it would open a path for a new and powerful technique for measuring very low fluxes of fast neutrons.

Table 2 – Measurements of U-238 and Th-232 in Boulby

| Source | Method | U-238 (ppb) | Th-232 (ppb) |
|---|---|---|---|
| Smith [34] | Ge gamma ray | 67±6 | 127±10 |
| Tziaferi [33] | Gd scintillator | 95±34(stat) ±21(sys) | 190±69(stat) ±42(sys) |
| This work | NITPC | 77±20(stat) ±7(sys) | 150±40(stat) ±10(sys) |

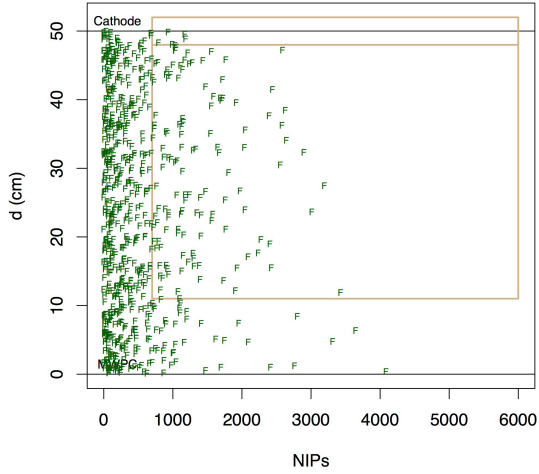

Fig. 10 – Simulated distribution of 700 F recoils from 100 GeV WIMPs in the *d* vs. *NIPs* space.

## 6. Discussion of Shielded Results and WIMP Limits

The calculation of dark matter spin-dependent limits proceeded as follows. For each of 32 WIMP masses, 9,000 fluorine recoils were simulated, using the parameters and equations found in [35] ($v_o$ = 230 km/s, $v_E$ = 244 km/s and $v_{esc}$ = 600 km/s). The recoils were distributed uniformly in *d*. An example of F recoils generated by 100 GeV WIMPs is shown in Fig. 10.

As before these MC data were binned in the usual way, multiplied by the efficiency map and the predicted number of detected events calculated for each WIMP mass. The ratio of the resulting summed numbers divided by 9,000 then allowed us to calculate the overall detection efficiency for each WIMP mass. The expected number of detected recoils for each WIMP mass was found by multiplying the nominal rate, calculated using the parameters and equations in [35] (with $\rho_D$ = 0.3 GeV/c$^2$/cm$^3$), by the exposure time (54.7 days) and by the detection efficiency for a fixed WIMP-nucleus cross-section. This cross-section was then scaled to produce 2.3 detected fluorine recoils to provide a 90% confidence limit cross-section. Finally, this was converted to a WIMP-proton spin-dependent cross section using the method of Tovey et al. [36].

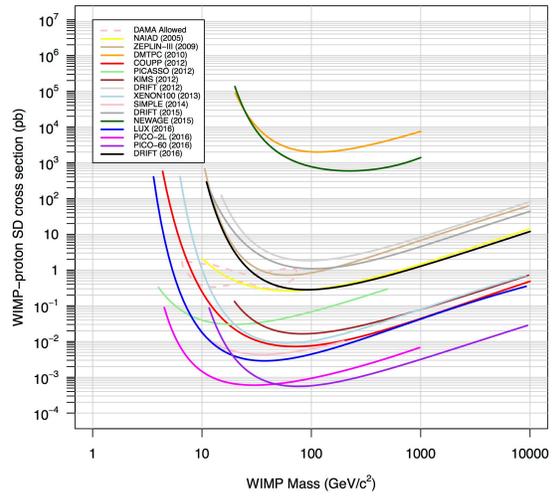

Fig. 11 – DRIFT's 90% C.L. upper limits on the WIMP-proton spin-dependent cross section as function of mass are shown in black. These limits exclude a portion of the DAMA allowed (3σ and no channeling [37]) region shown with a pink dashed curve. Also shown are the 90% C.L. from NAIAD [38], ZEPLIN-III [39], DMTPC [40], COUPP [41], PICASSO [42], KIMS [43], DRIFT[22,43], XENON100 [45], SIMPLE[46], NEWAGE [47], LUX [48], PICO-2L [49] and PICO-60 [50]. Note though that only DRIFT, DMTPC and NEWAGE have directional capabilities.

The resulting data were then smoothed to produce the exclusion curve shown in Fig. 11. These limits are approximately 4 times better than our high-threshold limits published previously [23]. A fraction of parameter space permitted by DAMA was excluded, a first for a directional detector [37].

The minimum excluded cross section is 0.28 pb at 100 GeV WIMP mass equivalent to the 2005 NAIAD NaI result and 3.5 orders of magnitude better than any other directional detector limit. The reason DRIFT's superior sensitivity is twofold and both have to do with our use of negative ions. First negative ion drift results in minimal diffusion which preserves directionality and gamma rejection at low energy. Thus DRIFT is able to operate at a lower threshold than other directional detectors. Second the use of negative ion drift allows for fiducialization to exclude RPR contamination.

## 6. Conclusion

DRIFT's spin-dependent limit has improved over the previous limit [23] by a factor of 4. The current limit is several orders of magnitude better than other directional limits. With the help of an improved gamma rejection factor, results from an unshielded run find candidate events consistent with detection of ambient fast neutrons from the walls of the Boulby cavern. A further improvement in the gamma rejection factor could yield a new technology capable of robust detection of low flux neutrons. DRIFT-IId continues to operate and will further explore the limits of low background nuclear recoil directional detection.

## Acknowledgements

We acknowledge the support of the US National Science Foundation (NSF). This material is based upon work supported by the NSF under Grant Nos. 1407754, 1103511, 1506237. J.B.R.B. acknowledges the support of the Alfred P. Sloan Foundation (BR2012-011), the National Science Foundation (PHY-1649966), the Research Corporation for Science Advancement (Award #23325) and the Massachusetts Space Grant Consortium (NNX16AH49H). We are grateful to Cleveland Potash Ltd and the Science and Technology Facilities Council (STFC) for operations support and use of the Boulby Underground Science Facility.

## References

[1] J. Beringer, et al., Phys. Rev. D 86 (2012) 010001. (Particle Data Group).
[2] J. L. Feng, Ann. Rev. Astron. Astrophys. (2010) 495.
[3] R. Bernabei, et al., Eur. Phys. J. C 67 (2010) 39.
[4] R. Agnese et al. (SuperCDMS Collaboration), Phys. Rev. Lett. 116 (2016) 071301.
[5] E. Aprile et al. (XENON100 Collaboration), Phys. Rev. Lett. 109 (2012) 181301.
[6] P. Agnes et al. (DarkSide Collaboration), Phys. Rev. D 93 (2016) 081101.
[7] Xiang Xiao et al. (PandaX Collaboration), Phys. Rev. D 92 (2015) 052004.
[8] D.S. Akerib et al. (LUX Collaboration),

Phys. Rev. Lett. 116 (2016) 161301.
[9] S. Ahlen, et al., Int. J. Mod. Phys. A 25 (01) (2010) 1.
[10] F. Mayet, et al., Phys. Rep. 627 (2016) 1.
[11] C. J. Copi, J. Heo, L. M. Krauss, Phys. Lett. B 461 (1999) 43.
[12] B. Morgan, A. Green, N. Spooner, Phys. Rev. D 71 (2005) 103507.
[13] A. Green, B. Morgan, Astropart. Phys. 27 (2-3) (2007) 142.
[14] D. Nygren, J. Phys. Conf. Ser. 460 (2013) 012006.
[15] P. Belli, et al., Nuovo Cim. C 15 C (1992) 475.
[16] F. Cappella, et al., Eur. Phys. J. C 73 (2013) 2276.
[17] N. J. C. Spooner, J. W. Roberts, D. R. Tovey, Proc. First Int. Work. Identif. Dark Matter, World Scientific (1997) 481.
[18] Y. Shimizu, et al., Nucl. Instr. Methods A 496 (2003) 347.
[19] Battat et al., Phys. Rept. 662, (2016) 1.
[20] D.P. Snowden-Ifft, C.J. Martoff, and J.M. Burwell, Phys. Rev. D 61 (2000) 1.
[21] D. P. Snowden-Ifft, J.-L. Gauvreau, Rev. Sci. Instrum. 84 (2013) 053304.
[22] J.B.R. Battat, et al., Nucl. Instr. Methods A 794 (2014) 33.
[23] J. B. R. Battat, et al., Phys. of the Dark Universe (2015) 1-7.
[24] D. Snowden-Ifft, Rev. Sci. Instrum. 85 (2014) 013303.
[25] J. B. R. Battat, et al., J. Inst. 10 (2016) 10019.
[26] G.J. Alner, et al., Nucl. Instr. Methods A 555 (2005) 173.
[27] S. Burgos, et al., Astropart. Phys. 28 (2007) 409.
[28] K. Pushkin and D.P. Snowden-Ifft, Nucl. Instr. Methods A 606 (2009) 569.
[29] S. Burgos, et al., Nucl. Instr. Methods A 600 (2009) 417.
[30] D. P. Snowden-Ifft, et al., Nucl. Instr. Methods A 498 (2003) 155.
[31] S. Agostinelli et al., Nucl. Instr. Methods A 506 (2003) 250.
[32] http://hepwww.rl.ac.uk/ukdmc/Radioactivity/rock.html
[34] P.F. Smith, et al., Astropart. Phys. 22 (2005) 409.
[33] E. Tziaferi, et al., Astropart. Phys. 27 (2007) 326.
[35] J.D. Lewin and P.F. Smith, Astropart. Phys. 6 (1996) 87.
[36] D.R. Tovey, et al., Phys. Lett B 488 (2000) 17.
[37] C. Savage et al., J. Cosmol. Astropart. Phys. 04 (2009) 010.
[38] G.J. Alner et al., Phys. Lett B 616 (2005) 17.
[39] V. Lebedenko et al., Phys. Rev. Lett. 103 (2009) 151302.
[40] S. Ahlen, et al., Phys. Lett. B, 695 (2011) 124.
[41] E. Behnke, et al., Phys. Rev. D 86 (2012) 052001.
[42] S. Archambault et al., Phys. Lett. B 711 (2012) 153.
[43] S. C. Kim et al., Phys. Rev. Lett. 108 (2012) 181301.
[44] E. Daw, et al., Astropart. Phys. 35 (2012) 397.
[45] E. Aprile et al., Phys. Rev. Lett. 111 (2013) 021301.
[46] M. Felizardo, et al., Phys. Rev. D 89 (2014) 072013.
[47] K. Nakamura, et al., Prog. Theo. and Exp. Phys. 4 (2015).
[48] Akerib, et al., Phys. Rev. Lett 16 (2016) 161302.
[49] C. Amole, et al., Phys. Rev. D 93 (2016) 061101.
[50] C. Amole et al., Phys. Rev. D 93 (2016) 052014.